\documentstyle[preprint,revtex]{aps}
\begin{document}
\begin{title}
Correlation functions in super Liouville theory
\end{title}
\author{E. Abdalla$^1$, M.C.B. Abdalla$^2$,}
\author{D. Dalmazi$^2$, Koji Harada$^{1(a)}$}
\begin{instit}
$^1$Instituto de F\'\i sica, Univ. S\~ao Paulo, CP 20516, S\~ao Paulo, Brazil

$^2$Instituto de F\'\i sica Te\'orica, UNESP, Rua Pamplona 145,

CEP 01405, S\~ao Paulo, Brazil
\end{instit}
\begin{abstract}
We calculate three- and four-point functions in super Liouville
theory coupled to super Coulomb gas on world sheets
with spherical topology.  We first integrate  over the zero mode and assume
that a parameter takes an integer value. After calculating the amplitudes, we
formally continue the parameter to an arbitrary real number. Remarkably the
result is completely parallel to the bosonic case, the amplitudes being
of  the same form as those of the bosonic case.
\end{abstract}
\newpage
\narrowtext
The matrix model definition  of 2D-gravity has been proving to be very
powerful in calculating correlation functions\cite{1}, although it seems
difficult to generalize the results to supersymmetric theories.
On the other hand, in the continuum approach (Liouville theory)\cite{2,3,4,5}
it is difficult to calculate correlation functions, while its supersymmetric
generalization (super Liouville theory)\cite{6,7,8} is well known.
Recently, however, several authors\cite{9,11,12,13,10} have exactly calculated
correlation functions in the continuum approach to conformal matter fields
coupled to 2D-gravity. (See also Ref. \cite{14}). They have used a technique
based on
the integration over the Liouville zero mode, and their results agree with
those obtained earlier in the discrete approach (matrix models). It is thus
very urgent to extend the continuum method\cite{15} to the supersymmetric case,
 i.e., superconformal matter fields coupled to 2D-supergravity.

The aim of this Letter is to calculate the three- and four-point functions in
super Liouville theory coupled to superconformal matter with the central charge
$\hat c<1$, represented as super Coulomb gas\cite{16}. Our approach is close to
that of Di Francesco and Kutasov\cite{11}. The result is remarkable and is
very parallel to the bosonic case;  it amounts to a
redefinition of the cosmological constant and of the primary superfields,
resulting  the same amplitudes as those of the bosonic theory.

The relevant framework has been given by Distler, Hlousek and Kawai\cite{8}.
With a translation invariant measure, the total action is given by $S=S_{SL}
+S_M$,
\begin{eqnarray}
S_{SL}&=&{1\over 4\pi}\int d^2{\bf z}\hat E\left( {1\over 2}\hat D_\alpha
\Phi_{SL}\hat D^\alpha \Phi_{SL}-Q\hat Y\Phi_{SL}-4i\mu
e^{\alpha_+\Phi_{SL}}\right)\quad ,\\
S_M &=&{1\over 4\pi}\int d^2{\bf z}\hat E({1\over 2}\hat D_\alpha \Phi_{M}\hat
D^\alpha \Phi_{M} + 2i\alpha_0\hat Y\Phi_M)\quad ,
\end{eqnarray}
where $\Phi_{SL}$, $\Phi_M$, are super Liouville and matter superfields
respectively. (See Refs.\cite{8,17}). The matter sector has the central charge
$\hat c_m=1-8\alpha_0^2$. The parameters $Q$ and $\alpha_\pm $ are given by
\begin{equation}
Q=2\sqrt {{1+\alpha_0^2}}\quad ,\quad
\alpha_\pm =-{Q\over 2}\pm {1\over 2}
\sqrt {Q^2-4}=-{Q\over 2}\pm \vert \alpha_0\vert \quad .
\end{equation}
The (gravitationally dressed) primary superfield $\widetilde \Psi_{NS}$ is
given by
\begin{equation}
\widetilde \Psi_{NS}({\bf z},k)=d^2{\bf z}\hat E e^{ik\Phi_M({\bf z})}e^
{\beta(k)\Phi_{SL}({\bf z})},
\end{equation}
\begin{equation}
\beta(k)=-{1\over 2}Q+\vert k-\alpha_0\vert\quad .\label{5}
\end{equation}
Note that the expressions for $\alpha_\pm $ and $\beta(k)$ are the same in
terms
of $Q$ and $\alpha_0$ as those of the bosonic theory.

 Screening charges in the matter sector are of the form
$d^2{\bf z}e^{id_\pm\Phi_M({\bf z})}$, where $d_\pm$ are the two solutions of
${1\over 2}
d(d-2\alpha_0)={1\over 2}$. In this Letter, however, we will concentrate on the
case without screening charges. The case with screening charges, $N(\ge 4)$-
point functions and the inclusion of the Ramond sector will be discussed
elsewhere.

We shall calculate three-point functions of the primary field $\widetilde
\Psi_{NS}$ on
world sheets with spherical topology (without screening charges), that is,
\begin{equation}
\left\langle\prod_{i=1}^3\int \widetilde \Psi_{NS}({\bf z}_i,k_i)\right\rangle
\equiv \int
[{\cal D}_{\hat E}\Phi_{SL}][{\cal D}_{\hat E} \Phi_M]\prod
_{i=1}^3\widetilde \Psi_{NS}({\bf z}_i,k_i)  e^{-S} \quad .
\end{equation}

Our first step is to integrate over the zero  modes,
\begin{eqnarray}
& &\left\langle\prod_{i=1}^3\int \widetilde \Psi_{NS}({\bf z}_i,k_i)\right
\rangle \equiv 2\pi
\delta \left( \sum _{i=1}^3k_i-2\alpha_0\right){\cal
A}(k_1,k_2,k_3),\nonumber\\
&&{\cal A}(k_1,k_2,k_3)\! =\!  \Gamma(-s)({-\pi\over 2})^3({i\mu
\over \pi})^s\! \left\langle \int \prod_{i=1}^3d^2{\bf \tilde z}_i
e^{ik_i\Phi_M(
\tilde {\bf z}_i)}e^{\beta_i\Phi_{SL}({\bf \tilde z}_i)}
\left(\! \int d^2{\bf z} e^{\alpha_+
\Phi_{SL} ({\bf z})}\right) ^s\!\right\rangle_0 \! ,\nonumber \\
\end{eqnarray}
where $\langle \cdots \rangle _0$ denotes the expectation value evaluated in
the free theory $(\mu =0)$ and we have absorbed the factor
$[\alpha_+(-\pi/2)^3]^{-1}$ into the normalization of the path integral.
The parameter $s$ is defined as
\begin{equation}
s=-{1\over \alpha_+}\left[ Q+\sum_{i=1}^3\beta_i\right] \quad .
\label{8}\end{equation}

In general, $s$ can take any real value and there is no obvious way of
calculating the path-integral. However, if we assume that $s$ is a non-negative
integer\cite{8,9,11,12,13,10}, this is just a free-field
correlator. Under this assumption, we evaluate the path-integral,
and formally continue $s$ to
non-integer values. For $s$ non-negative integer,
\begin{eqnarray}
&&{\cal A}(k_1,k_2,k_3)=\Gamma(-s)({-\pi\over 2})^3\left( {i\mu\over
\pi}\right) ^s \nonumber\\
&\times&\int \prod_{i=1}^3d^2{\bf \tilde z}_i\prod_{i=1}^sd^2{\bf z}_i
\prod_{i<j}^3\vert {\bf \tilde
z}_{ij}\vert ^{2k_ik_j-2\beta_i\beta_j}\prod_{i=1}^3\prod_{j=1}^s\vert
 \tilde z_i-z_j-\tilde \theta _i\theta_j
\vert ^{-2\alpha_i\beta_i}\prod_{i<j}^s\vert
{\bf z}_{ij}\vert ^{-2\alpha_+^2}\nonumber \\
&=&\Gamma(-s)({-\pi\over 2})^3\left( {i\mu\over \pi}\right) ^s\int \prod_{i=1}
^sd^2 {\bf z}_id^2\tilde \theta \prod_{i=1}^s\vert z_i+\tilde \theta
\theta _i\vert
^{-2\alpha_+\beta _1}\prod_{1=1}^s\vert 1-z_i\vert ^{-2\alpha_+\beta_2}\prod
_{i<j}^s\vert {\bf z}_{ij}\vert ^{-2\alpha_+^2}\nonumber\\
\end{eqnarray}
We have divided by the $\widehat{SL_2}$ volume by setting $\tilde z_1=0, \tilde
 z_2=1,\tilde z_3=\infty, \tilde \theta_2=\tilde \theta_3=0$ and $\tilde \theta
_1\equiv \tilde \theta $. The integral is a supersymmetric generalization of
(B.9) of Ref.\cite{18}.
Alternatively, using  $\Phi_{SL}=\phi +\theta\psi +\bar \theta \bar
\psi$, we can write
\begin{eqnarray}
&&{\cal A}(k_1,k_2,k_3)\nonumber\\
&&=\Gamma(-s)({-\pi\over 2})^3\left( {i\alpha_+^2\mu \over
\pi}\right) ^s\beta_1^2
\int \prod_{i=1}^sd^2z_i\prod_{i=1}^s\vert z_i\vert ^{-2\alpha_+
\beta_1} \vert 1- z_i\vert ^{-2\alpha_+\beta_2}\prod _{i<j}^s\vert z_i-z_j\vert
^{-2\alpha_+^2}\nonumber\\
&&\phantom{ 123456789012345} \times \left\langle \overline \psi\psi(0)\overline
\psi\psi(z_1)\cdots\overline \psi\psi(z_s)\right\rangle_0 \quad .
\end{eqnarray}
This is non-vanishing only for $s$ odd; we thus write
 $s\equiv 2m+1$.
One may evaluate $\langle \overline \psi \cdots \overline \psi \rangle _0$ and
$\langle \psi \cdots \psi \rangle_0$ independently. Since the rest of the
integrand is symmetric, one may write the result in a simple form by
relabelling
coordinates:
\begin{equation}
{\cal A}(k_1,k_2,k_3)=
-i({-\pi\over 2})^3\Gamma(-s)\Gamma(s+1){1\over \alpha_+^2}\left(
{\alpha_+^2\mu \over\pi}\right) ^s I^m(\alpha,\beta;\rho)
\end{equation}
where
\begin{eqnarray}
&&I^m(\alpha,\beta;\rho) = {\alpha ^2\over 2^mm!}\! \int \! d^2w \!
\prod_{i=1}^m \! d^2\zeta_id^2\eta_i\vert w\vert ^{2\alpha-2}\vert 1\! - \!
w\vert ^{2\beta}
\prod_{i=1}^m\! \vert w\! -\! \zeta_i\vert ^{4\rho}\vert w\! -\! \eta_i
\vert ^{4\rho }\nonumber\\
&&\times \prod_{i=1}^m\vert \zeta_i\vert ^{2\alpha} \vert \eta_i\vert ^
{2\alpha}\vert
1-\zeta_i\vert ^{2\beta}
\vert 1-\eta_i \vert ^{2\beta} \prod _{i,j}^m \vert \zeta
_i-\eta_j\vert ^{4\rho} \prod _{i<j}^m\vert \zeta_i-\zeta_j\vert ^{4\rho} \vert
\eta_i-\eta_j\vert ^{4\rho}\prod _{i=1}^m\vert \zeta_i-\eta_i\vert
^{-2}\nonumber\\
\end{eqnarray}
and $\alpha=-\alpha_+\beta_1 $, $\;\beta =-\alpha_+\beta_2$, $\; \rho=-{1\over
2}\alpha_+^2$.

We now have to calculate $I^m(\alpha,\beta; \rho)$. First of all, we observe
that $I^m(\alpha,\beta; \rho)$ should be symmetric in $\alpha $ and $\beta $:
$I^m(\alpha,\beta; \rho) = I^m(\beta,\alpha; \rho)$. It is easy to check that
when  $m=0$. The large-$\alpha$ and large-$\beta$
behaviors are consistent with  it, and it  is physically natural
because the amplitude should be symmetric under the exchange of two external
momenta. Thus $I^m(\alpha,\beta; \rho)$ exhibits the
following symmetry
\begin{equation}
I^m(\alpha,\beta;\rho)=I^m(-1-\alpha-\beta-m\rho,\beta;\rho) \label{13}
\end{equation}
Thus we may write $I^m(\alpha,\beta;\rho)$ in the following way
\begin{eqnarray}
&&I^m(\alpha,\beta;\rho)=C_m(\alpha,\beta;\rho)\prod_{i=0}^m\Delta( 1\! +\!
\alpha \! +\! 2i\rho) \Delta( 1\! +\! \beta \! +\! 2i
\rho) \Delta(\!  -\! \alpha \! - \! \beta \! +\! (2i \! -\! 4m)\rho)\nonumber\\
&\times&\prod_{i=1}^m\Delta( {1\over 2}+\alpha+(2i-1
)\rho) \Delta( {1\over 2}+\beta+(2i-1)
\rho) \Delta( -{1\over 2}-\alpha-\beta+(2i
-4m-1)\rho) \nonumber\\
\end{eqnarray}
where $C_m(\alpha,\beta;\rho)$ has the same symmetries as
$I^m(\alpha,\beta;\rho)$,
and where $\Delta(x)\equiv
\Gamma(x)/\Gamma(1-x)$. By looking at the large-$\alpha$ behavior:
\begin{equation}
I^m(\alpha,\beta;\rho)\sim \alpha^{-2m-2(2m+1)\beta-4\rho m (2m+1)}\quad,
\end{equation}
one can confirm that $C_m(\alpha,\beta;\rho)$ is, as a function of $\alpha$,
bounded as $\vert \alpha\vert \to \infty$ and analytic. This means that $
C_m(\alpha,\beta;\rho)$ is independent  of $\alpha$, and by symmetry, of
$\beta$ as well; $C_m=C_m(\rho)$.

It is hard to calculate $C_m(\rho)$. For this purpose, it is useful to consider
the simpler integral:
\begin{eqnarray}
J^m(\alpha,\beta;\gamma;\rho)&=& \int \prod_{i=1}^md^2\zeta_id^2\eta_i
\prod_{i=1}^m\vert \zeta_i\vert ^{2\alpha} \vert \eta_i\vert ^{2\alpha}\vert
1-\zeta_i\vert ^{2\beta}
\vert 1-\eta_i \vert ^{2\beta} \prod _{i,j}^m \vert \zeta
_i-\eta_j\vert ^{4\rho} \nonumber\\
&&\times \prod _{i<j}^m\vert \zeta_i-\zeta_j\vert ^{4\rho} \vert
\eta_i-\eta_j\vert ^{4\rho}
\prod _{i=1}^m\vert \zeta_i-\eta_i\vert ^{4\gamma}\quad .
\end{eqnarray}
By using similar arguments, one may obtain
\begin{eqnarray}
&&J^m(\alpha,\beta;\gamma;\rho)\nonumber \\
&=&\widetilde C_m(\gamma;\rho)\prod_{i=0}^{m-1}
\Delta( 1+\alpha+2i\rho) \Delta( 1+\beta
+2i\rho) \Delta( -1-\alpha-\beta-
2\gamma+(2i-4m+2)\rho) \nonumber\\
&\times&\!\prod_{i=1}^m\!\Delta(\! 1\!+\!\alpha\!+\!\gamma\!+\!
(2i\!-\!1)\rho) \Delta( 1\!+\!\beta\!+\!\gamma\!+
(2i\!-\!1)\rho) \Delta( -1\!-\!\alpha\!-\!\beta\!-
\!\gamma\!+(2i\!-\!4m\!+\!2)\rho). \nonumber\\
\end{eqnarray}
Again, it is very difficult to calculate $\widetilde C_m(\gamma;\rho)$.
Unfortunately we could not get it in a rigorous way. A series of trials
and errors, however, led us to the following form;
\begin{equation}
\widetilde C_m(\gamma;\rho)={\pi^{2m}\over 2^m}m! \left[ \Delta\left( -(\gamma
+\rho) \right) \right]^{2m}\prod_{i=1}^m\Delta \left(
1+2(\gamma+i\rho)\right) \Delta \left( 1+\gamma+(2i-1)\rho\right) .
\label{18}\end{equation}
This is consistent with $\widetilde C_1(\gamma;\rho)={\pi^{2}\over 2}
\left[ \Delta\left( -(\gamma +\rho)\right) \right]^{2}\Delta \left(
1+\gamma+\rho\right) \Delta \left( 1+2(\gamma+\rho)\right)$
and the two other (calculable) cases $\rho=0$ and $\gamma=0$ (up to
 symmetry factors). It is very difficult to get anything
else consistent with these
constraints. And {\it a posteriori } it seems to be correct since it gives a
physically reasonable result. Let us assume that (\ref{18}) is correct and
see its consequences.

The two integrals are related by
\begin{equation}
I^m(\epsilon,\beta;\rho)=-{\pi\over 2^m m!}\Delta\left( 1+\epsilon
\right) \Delta\left( 1+ \beta\right) \Delta\left( -\epsilon
-\beta\right) J^m(2\rho,\beta;-1/2;\rho)\; .
\end{equation}
Therefore $C_m(\rho)=-(\pi/2^mm!) \widetilde C_m(-1/2,\rho)\Delta({1\over
2}-\rho)\Delta({1\over 2}+(2m+1)\rho)$. If we substitute (\ref{18}) we
get
\begin{equation}
C_m(\rho)=-{\pi^{2m+1}\over 2^{2m}}\left[ \Delta\left( {1\over
2}-\rho\right) \right]^{2m+1}\prod_{i=1}^m\Delta\left( 2i\rho
\right) \prod _{i=0}^m\Delta\left( {1\over 2}+(2i+1)\rho\right) \quad .
\end{equation}

Now we are ready to write down the amplitude. Without loss of generality, we
can choose $k_1,k_3\ge\alpha_0, k_2\le \alpha_0$. By using (\ref{5}),
(\ref{8}) and $\sum_{i=1}^3 k_i=2\alpha_0$, one gets
\begin{eqnarray}
&&\beta=\left\{
\begin{array}{l} \rho (1-s)\; \; (\alpha_0>0)\nonumber \\
-{1\over 2}-\rho s \; \; (\alpha_0<0).\nonumber \\
\end{array}\right. \nonumber
\end{eqnarray}

It is easily seen that, for $\alpha_0>0$, ${\cal A}=0$ identically, as in the
bosonic theory. For $\alpha_0<0$, there are many cancellations, leading to
\begin{eqnarray}
&&I^m(\alpha,\beta,\rho)\nonumber \\
&=&C_m(\rho)\prod_{i=0}^m\Delta\left( {1\over 2}-(2i+1)
\rho\right) \prod_{i=1}^m\Delta\left( -2i\rho \right)
 \Delta \left( 1+\alpha+2m\rho\right) \Delta\left(
{1\over 2}-\alpha+\rho\right) \nonumber\\
&=&(-1)^{m+1}{\pi^{2m+1}\over (m!)^2}\left[ \Delta \left( {1\over 2}-\rho
\right) \right]^{2m+1}(4\rho)^{-2m}\Delta \left( 1+\alpha+2m
\rho\right) \Delta\left( {1\over 2}-\alpha + \rho\right)
\, .
\end{eqnarray}
 We finally obtain the three-point function:
\begin{eqnarray}
{\cal A}(k_1,k_2,k_3)&=& ({-i\pi\over 2})^3\left[ {\mu\over 2}\Delta\left(
{1\over 2}-\rho\right) \right]^s\Delta\left( {1\over 2}-{s\over 2}
\right)\Delta \left( 1+\alpha+2m\rho\right) \Delta\left(
{1\over 2}-\alpha+\rho\right)\nonumber\\
&=&\left[ {\mu\over 2}\Delta\left( {1\over 2}-\rho
\right)\right]^s\prod_{j=1}^3(-{i\pi\over 2})\Delta\left( {1\over 2}[1+
\beta_j^2-k_j^2]\right)\quad .
\end{eqnarray}

 By redefining the cosmological constant and the primary superfield
$\widetilde\Psi_{NS}$ as
\begin{equation}
\mu  \to  {2\over \Delta \left( {1\over 2}-\rho\right)}\mu \quad
,\quad
\widetilde\Psi_{NS}(k_j)\to {1\over (-{i\over 2}\pi)\Delta
\left( {1\over 2}[1+\beta_j^2-k_j^2]\right)}\widetilde\Psi_{NS}(k_j)\quad ,
\end{equation}
we get our main result
\begin{equation}
{\cal A}(k_1,k_2,k_3)=\mu ^s\quad .
\end{equation}
Remarkably, this amplitude is of the same form as the bosonic one\cite{11}.

It is natural to expect that this  feature continues to be true for
$N (\ge 4)$- point functions.
In fact, for $k_1, k_2, k_3 \ge \alpha_0,\; k_4\le \alpha_0<0$ (and without
screening charges), the four point function turns out to be
\begin{equation}
{\cal A}(k_1,k_2,k_3,k_4)=(s+1)\mu ^s \; ,
\end{equation}
with the same redefinition of the cosmological constant and the primary
superfields. In order to get the amplitude for general $k_i$, one may argue, as
in Ref. \cite{11}, that non-analyticity comes entirely from massless
intermediate states and one may calculate the amplitude by
using the analyticity of the one
particle irreducible (1PI) correlators. After setting $\mu =1$, we obtain the
four-point function for all $k_i$:
\begin{equation}
{\cal A}(k_1, k_2, k_3,  k_4)=-\alpha_-[\vert k_1\! +\! k_2\! -\!
\alpha_0\vert\!  +\! \vert
k_1\! +\! k_3\! -\!\alpha_0\vert\! +\!\vert k_1\! +\!
k_4\! -\! \alpha_0\vert ]\! +\! {\cal A}_{1PI} ,
\end{equation}
with ${\cal A}_{1PI}=-{1\over 2}(1+\alpha_-^2)$. Compare with Eq. (37) in Ref.
\cite{11}. The analogy to the bosonic case is obvious.  A detailed account
will appear elsewhere.

The close connection to the bosonic amplitudes was suggested from super-KP
systems\cite{19}, and more recently, from supermatrix models\cite{20}.
We think, however, that our demonstration is more direct.

In conclusion, we calculated the three- and four-point
functions of super Liouville theory
coupled to super Coulomb gas (without screening charges) on a sphere and found
that they are  essentially the same as those of the usual
Liouville theory, obtained
in Ref. [11]. As a by-product we get the supersymmetric generalization of (B.9)
formula of Ref. \cite{18} ($s=2m+1$),
\begin{eqnarray}
&&{1\over s!}
\int \prod_{i=1}^sd^2{\bf z}_id^2\tilde \theta \prod_{i=1}^s\vert z_i+
\tilde \theta \theta_i\vert ^{2\alpha}\prod_{i=1}^s\vert 1-z_i\vert
^{2\beta}\prod_{i<j}^s\vert {\bf z}_{ij}\vert ^{4\rho}\nonumber\\
&=& (-1)^m\pi^{2m+1} \rho^{2m} \Delta\left( {1\over 2}-\rho\right)^{2m+1}
\prod _{i=1}^m\Delta(2i\rho)
\prod_{i=0}^m\Delta\left( {1\over 2}+(2i+1)\rho\right)\nonumber\\
&\times &
\prod_{i=0}^m\Delta(1+\alpha+2i\rho)\Delta(1+\beta+2i\rho)\Delta(-\alpha-\beta+
(2i-4m)\rho)\nonumber\\
&\times& \prod_{i=1}^m\!\Delta\!\left( {1\over 2}+\!\alpha\! +(2i\! -\!
1)\rho\right)\! \Delta\!
\left( {1\over 2}+\!\beta\! +(2i\! -\! 1)\rho\right)\! \Delta\!
 \left( -{1\over 2}-\!\alpha\! -\!\beta\!
+(2i\! -\! 4m\! -\! 1)\rho\right)\, . \nonumber\\
\end{eqnarray}

\acknowledgments

K.H. would like to thank the members of Instituto de F\'\i sica Te\'orica,
UNESP, for their hospitality extended to him and FAPESP (\# 90/1799-9) for
the financial support. He also acknowledges communications on 2D
gravity with Y.Tanii and with N.Ishibashi. D.D thanks FAPESP (\# 90/2246-3)
for financial support.  Work of E.A. and M.C.B.A. is partially
supported by CNPq.

(a) Address after October 1991, Department of Physics, Kyushu University,
Fukuoka 812, Japan.

\end{document}